%% file: Dmix_ichep04_v1.6.tex
\documentclass[aps,prl,preprint,tightenlines,superscriptaddress,showpacs,byrevtex]{revtex4}
\usepackage{graphicx} % Include figure files
\usepackage{dcolumn}  % Align table columns on decimal point

\graphicspath{{ps}}

\begin{document}

%\vspace*{-3\baselineskip}
%\resizebox{!}{3cm}{\includegraphics{belle.eps}}

\preprint{\vbox{ \hbox{   }
                 \hbox{BELLE-CONF-0455}
                 \hbox{ICHEP04 11-0703} 
%                 \hbox{hep-ex nnnn, if available}
}}

\title{ \quad\\[0.5cm] Search for $D^0-\overline{D^0}$ mixing using semileptonic decays at Belle }

%%%% >>>>> insert the authorlist here. BEFORE the abstract !!!!! <<<<<

%\include{author-conf2004}
%\input{author}
\input{author-conf2004}

\begin{abstract}
A search for mixing in the neutral $D$ meson system was performed using 
semileptonic $D^0\to K^-e^+\nu$ decays. The flavor of neutral $D$
mesons at production was tagged by the charge of the slow pion from
the decay $D^{\ast+}\to D^0\pi^+$. The measurement was performed using
140~fb$^{-1}$ of data recorded by the Belle detector. 
The yield of
right and wrong sign decays arising from un-mixed and mixed
events, respectively, 
was determined by a fit to kinematic observables.
From the number
of signal events we derive an upper limit for the time-integrated 
mixing rate $r_D < 1.4\times 10^{-3}$ at $90\%$ C.L.
\end{abstract}

\pacs{13.20.Fc, 14.40.Lb}

\maketitle

%%%% >>>> keep the final version single-spaced
%\tighten
% -> Class revtex4 Warning: Command \tighten is obsolete; Invoke option tightenlines instead..

{\renewcommand{\thefootnote}{\fnsymbol{footnote}}}
\setcounter{footnote}{0}

\section{I Introduction}
\label{sec1}

\begin{figure}[htb]
\includegraphics[width=0.6\textwidth]{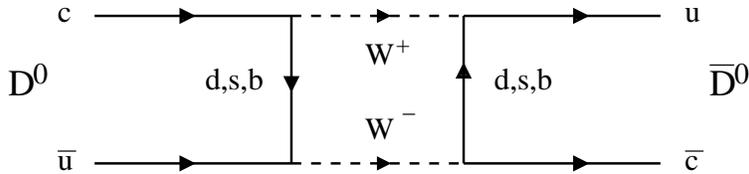}
\caption{Box diagram for the $D^0$ mixing amplitude.}
\label{fig_1}
\end{figure}

By considering the quark box diagrams describing
$D^0-\overline{D}^0$ mixing (Fig.\ref{fig_1}), it is easy to see why the
mixing rate in the charm sector is expected to be small. 
The amplitude
for these processes can be written as 
\begin{equation}
<\overline{D}^0|H_{\rm weak}|D^0>~\propto\sum_{i,j=d,s,b}
V_{ci}^\ast V_{ui}V_{cj}V_{uj}^\ast~{\cal{K}}(m_i^2,m_j^2)~~,
\label{eq1}
\end{equation}
where $V_{kl}$ are the Cabibbo-Kobayashi-Maskawa (CKM) matrix elements
\cite{CKM} and ${\cal{K}}$ is the loop function depending on the
masses of down-like quarks. The contribution of $b$ quarks is
negligible due to the small magnitude of $V_{cb}^\ast V_{ub}$. In the
limit that the $d$ and $s$ quark masses are equal, the unitarity of
the CKM matrix ensures a vanishing amplitude. However, the mass
difference of $d$ and $s$ quarks is
insufficient to result in a large mixing rate. Expectations for the
ratio of mixed to unmixed decay rates from the box diagram are of
the order of $10^{-10}-10^{-9}$ \cite{box_th}. However, due to
the smallness of the short distance contributions, long distance effects
could be important. The resulting predictions for the mixing probability depend on the 
treatment of these effects and cover a large range, 
reaching values as high as $10^{-3}$
\cite{long_th}. 

In direct searches, the signature of mixing is neutral $D$ meson
decay products with a charge combination opposite to that expected for a
produced flavor eigenstate. Examples of appropriate $D^0 \to f$ decay modes for
these 
methods are $D^0\to K^-\pi^+$ and $D^0\to
K^-\ell^+\nu$\footnote{Throughout the paper charge conjugated modes
are implied unless stated otherwise.}. If in the decay of an initially produced
$D^0$ a $K^+\pi^-$ or $K^+\ell^-\overline{\nu}$ 
final state ($\overline{f}$) is
observed, this represents an indication of a mixing transition 
$D^0\to\overline{D}^0\to 
\overline{f}$. 
While the semileptonic decay modes are more
difficult to reconstruct due to the presence of a neutrino in the final
state, the hadronic modes suffer from an irreducible background of
doubly Cabibbo suppressed (DCS) decays.

To search for $D^0-\overline{D^0}$ mixing we used the  
$D^{\ast+}\to D^0\pi_{\rm s}^+,~~D^0\to \overline{D^0} \to K^+e^-\overline\nu$ decay chain. The
flavor of the produced $D$ meson is tagged by the charge of the slow
pion ($\pi_{\rm s}^\pm$) from the $D^\ast$ decay. The charge of
the electron determines the
charm quantum number of the $D$ meson at decay. If all the final state
particles are reconstructed, signal events yield a sharp peak in the
$\Delta m = m(\pi_{\rm s}^+K^-e^+\nu)-m(K^-e^+\nu)$ distribution,
where $m(\pi_{\rm s}^+K^-e^+\nu)$ and $m(K^-e^+\nu)$ are the invariant
masses of the selected particles. In order to use the $\Delta m$
distribution to isolate the signal events, one needs to reconstruct the
momentum of the neutrino. Since the
kinematics of mixed and un-mixed decays are the same, both can be
selected using the $\Delta m$ observable.

Un-mixed decays, $D^{\ast+}\to D^0\pi_{\rm s}^+,~~D^0\to K^-e^+\nu$, 
result in the ``right sign" (RS) final state charge combination,
$\pi_{\rm s}^+K^-e^+$. Mixed decays, $D^{\ast+}\to D^0\pi_{\rm
s}^+,~~D^0\to \overline{D}^0 \to K^+e^-\overline{\nu}$, produce a
``wrong sign" (WS) charge combination, $\pi_{\rm s}^+K^+e^-$. Table
\ref{tab1} lists possible charge combinations of final state
particles used in the analysis. 
\begin{table}[h]
\begin{tabular}{|c|c|c|}
\hline\hline
Decay & Final state & Notation \\ \hline\hline
$D^{\ast+}\to D^0\pi_{\rm s}^+,~~D^0\to K^-e^+\nu$, 
un-mixed & $\pi_{\rm s}^+K^-e^+$ & Right sign (RS) \\ \hline
$D^{\ast+}\to D^0\pi_{\rm s}^+,~~D^0\to \overline{D}^0 \to
K^+e^-\overline{\nu}$, 
mixed    & $\pi_{\rm s}^+K^+e^-$ & Wrong sign (WS) \\ \hline
combinatorial background & $\pi_{\rm s}^+K^\pm e^\pm$ & Combinatorial
sign (CS) \\ \hline\hline
\end{tabular}
\caption{Summary of charge combinations and notations used in the
analysis.}
\label{tab1}
\end{table}
%\end{tabular}

In the following section we describe the selection of $D$ meson
semileptonic decay candidates. Section 3
%\ref{sec.3} 
describes the 
neutrino reconstruction and calculation of $\Delta m$. A background
description is presented in section 4
%\ref{sec.4} 
and the proper decay
time distribution of mixed and un-mixed decays is dealt with in
section 5.
%\ref{sec.5}. 
The succeeding section describes the fit of $\Delta m$
distributions, the obtained signal yield and the time integrated mixing rate $r_D$. 
In section 7
%\ref{sec.7}
we evaluate the systematic errors and describe several cross-checks of the 
measurement.

\section{II Selection of semileptonic $D$ decays}
%\label{sec.2}
%\setcounter{figure}{0}

\subsection{Data set}
For the present analysis, data collected by the Belle detector at
the center-of-mass energy of the $\Upsilon(4S)$ resonance, 
corresponding to an integrated luminosity of
140~fb$^{-1}$ were used. A detailed description of the Belle detector
can be found in \cite{Belle_det}. Simulated events were 
generated by the QQ generator and
processed with a full simulation of the Belle detector, using the
GEANT package \cite{qq98}. 
About 410~fb$^{-1}$ of $c\overline{c}$ quark pair Monte Carlo (MC) events
were used, together with about 164~fb$^{-1}$ light quark and about 100~fb$^{-1}$
$B\overline{B}$ meson pair MC events, which were rescaled by the
appropriate factors.

To simulate mixed $D$ meson decays, a
generic (non-mixed) MC was used with an appropriate re-weighting of
the proper decay time distribution. 

\subsection{Selection criteria} 

Electron-positron interactions with hadrons in the final state are
selected from data, based on the neutral cluster energy, total visible
energy in the center-of-mass system (cms), 
z component of total cms momentum and the position of reconstructed event
vertex \cite{hadron_B}, with an efficiency above 99\%.

The selection criteria were optimized using the simulated sample of
generic $u$, $d$, $s$, charm and $B$ meson pair events, by maximizing the
efficiency for non-mixed signal events multiplied by the purity of
the selected sample. Since the kinematic properties of mixed and
non-mixed events are the same, such an optimization ensures optimal
selection for the mixed events as well. 

For $\pi_s^\pm$ candidates, tracks with impact parameter with respect
to the interaction point in the radial
direction $|\delta r|<1$~cm and in the beam direction $|\delta z|<2$~cm
were considered. This procedure suppressed badly reconstructed tracks and tracks which do not
originate from the interaction point. A slow pion candidate was required to have a momentum of less
than 
600~MeV/$c$. To reduce the background from electrons misidentified as slow pion
candidates, we required the electron identification likelihood
\cite{e_id} for each track, based on
the information from the central drift chamber, aerogel
Cherenkov counter  and electromagnetic calorimeter, to be lower than
0.1. 
%The MC efficiency for reconstructing the $\pi_s^\pm$ from
%the signal was found to be around 
%$58\%$. 
%$60\%$.
To illustrate the $\pi_s^\pm$ selection criteria the MC efficiency for
reconstruction of a slow pion track was found to be around 60\%. 
Electron candidates were required to have a momentum
$p>600$~MeV/$c$ and an electron identification likelihood
\cite{e_id} above $0.95$. 
%The efficiency for the reconstruction of 
%a signal electron was found from simulated events to be approximately 
%43\%.
%$45\%$. 
By that 45\% of all simulated signal electrons are reconstructed.
From the remaining tracks in an event, 
 charged kaon candidates were chosen with momentum above
800~MeV/$c$. Using the information from the time of flight counters,
central drift chamber and aerogel 
Cherenkov counters, a relative likelihood $\cal{L}$ for a given track to be a
$K^\pm$ or $\pi^\pm$ is obtained \cite{hadron_B}. Kaon 
candidates were selected using 
${\cal{L}}(K^\pm)/({\cal{L}}(K^\pm)+{\cal{L}}(\pi^\pm))>0.5$. 
Using this selection
%$52\%$
about $50\%$ of signal kaons were retained.

In order to reduce the combinatorial background and the background from $B\overline{B}$ 
events, and to improve the resolution in the $\Delta m$, a requirement for the 
cms momentum of the kaon-electron
system, $p^\ast(Ke)>2$~GeV/$c$ was imposed. To further suppress $D^0$ mesons arising 
from $B\overline{B}$ events, we required the ratio of
the second and zeroth Fox-Wolfram moments $R_2$ to be greater than 0.2 \cite{fox-wolfram}.

Background from $D^0 \to K^- \pi^+$, with $\pi^+$ misidentified
as $e^+$ (RS sample) or with $K^-$ misidentified as
$e^-$ and $\pi^+$ as $K^+$ (WS sample),
was suppressed using a requirement on the invariant mass of the $K-e$ system,
$M(Ke) < 1.82$~GeV/$c^2$.
 Another source of background contributing 
to the RS and WS samples is the 
$D^0\to K^-K^+$ decay with either of the charged kaons misidentified
as an electron. This background is effectively
suppressed by the requirement $|M(KK)-m_{D^0}|>10$~MeV/$c^2$,  
where $M(KK)$ is the invariant mass of the kaon and electron candidate, 
calculated using the kaon
mass for both tracks.

An important source of background are the electrons from photon conversions,
which could be selected as electron or slow pion
candidates. 
Combinations where both $\pi^+_s$ and $e^-$ candidates are from
conversions, are almost
completely suppressed by requiring $M(e^+e^-) < 150$~MeV/$c^2$, where
the electron mass is assigned to the $\pi_s^+$ candidate. 
A further search for an additional $e^\pm$ of opposite charge
as the semileptonic electron candidate or as the slow pion candidate
was performed in each event.
%Such combinations could arise by considering an electron from
%the $\gamma$ conversion either as a signal electron or as a slow pion
%candidate.
Again, events with $M(e^+e^-)$
below 150~MeV/$c^2$ were rejected. 
For the events satisfying all the above criteria, the neutrino reconstruction
was performed. 

\section{III Neutrino reconstruction}
%\label{sec.3}
%\setcounter{figure}{0}

\begin{figure}[htb]
\includegraphics[width=0.4\textwidth]{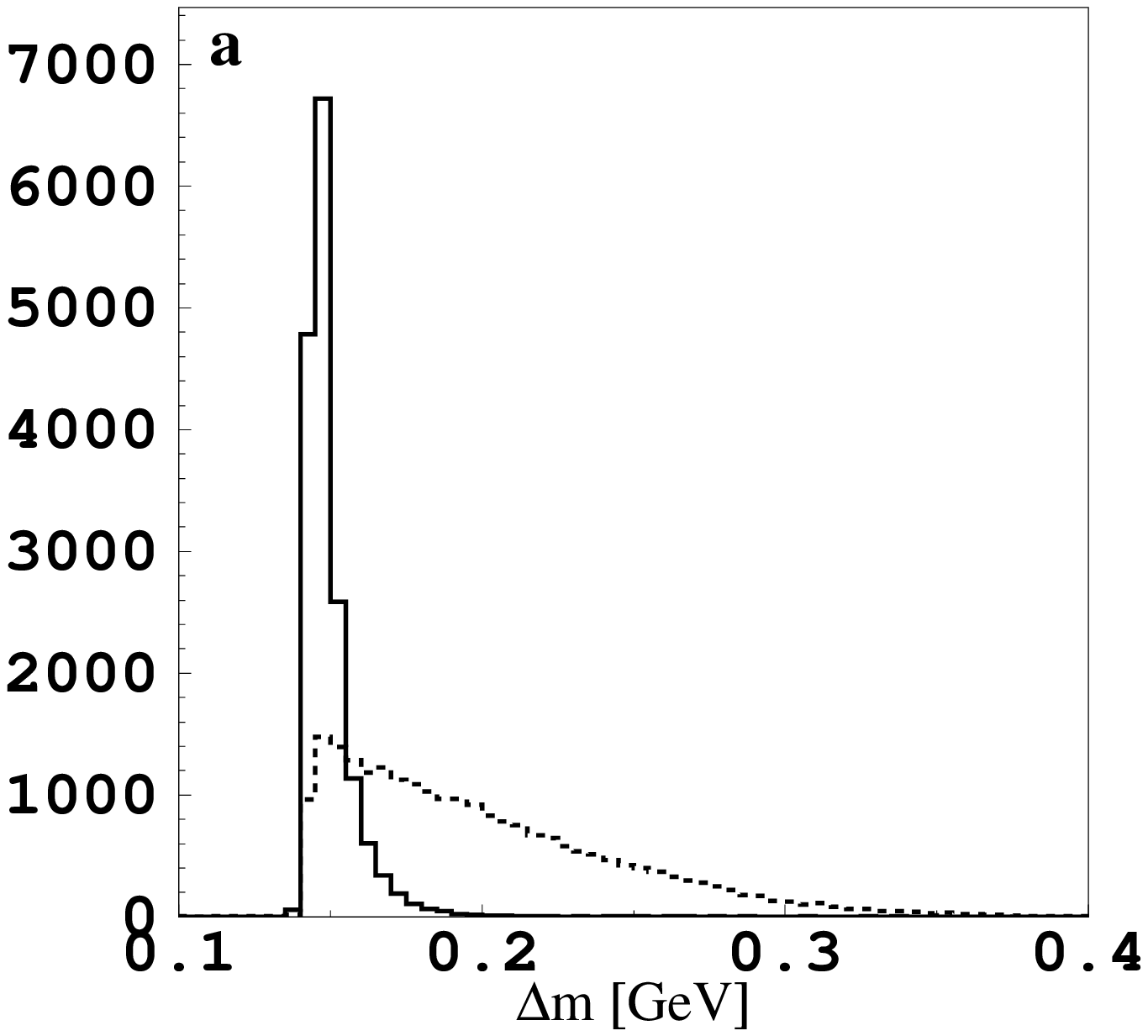}
\includegraphics[width=0.4\textwidth]{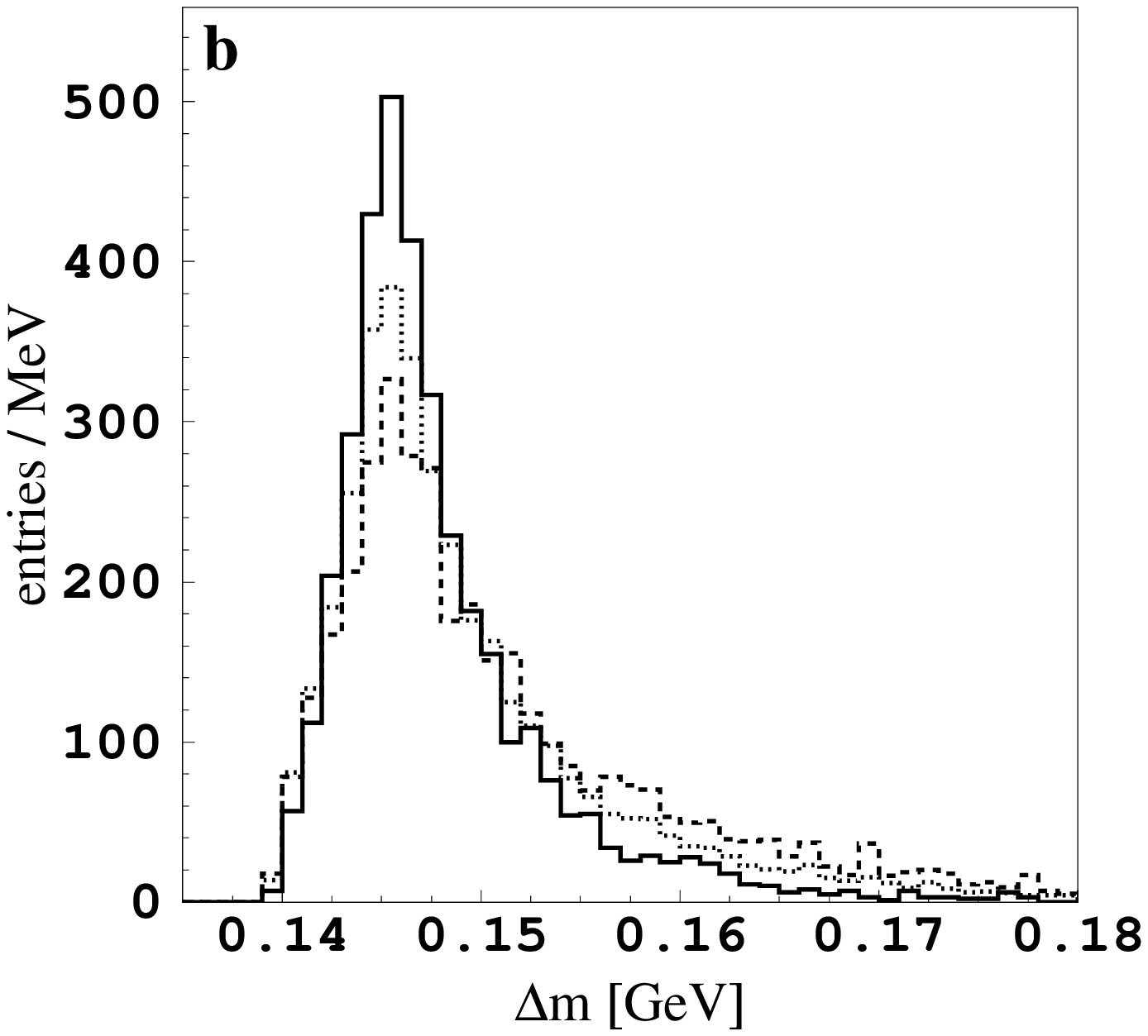}
\vspace*{-0.5cm}
\caption{a) $\Delta m$ distribution for simulated signal
events using the missing momentum for neutrino reconstruction (dashed
histogram) or the missing momentum together with the constraints on
the missing mass and the mass of the $D^{\ast +}$ (solid histogram) as described
in the text.  b) $\Delta m$ distribution for simulated signal
events in different bins of the momentum of the kaon-electron
system: $p^*(Ke) ~< ~2.0$~GeV/$c$ (dashed histogram), 2.0~GeV/$c~ <
p^*(Ke) < 3.2$~GeV/$c$ 
(dotted histogram), $3.2$~GeV/$c$~$ < ~p^*(Ke)$ (solid histogram). The
histograms are normalized to the same area.}
\label{fig_5}
\end{figure}

Four-momentum conservation in an $e^+e^-$ collision implies 
\begin{equation}
P_\nu=P_{\rm cms}-P_{\pi_s K e} -P_{\rm rest}
\label{eq4}
\end{equation}
for the signal decay, where cms stands for the center-of-mass
four-momentum of the $e^+e^-$ system, and the index $rest$ indicates the
four-momentum of all detected particles except the slow pion, charged
kaon and the electron candidate\footnote{In the following, $P$ denotes the 
particle's 4-momentum while 
$\vec{p}$ and $p$ denote the 3-dimensional momentum and its magnitude,
respectively.}. The variable
$P_{\rm rest}$ is calculated using all the 
remaining charged
tracks in the event with $|\delta r|<1$~cm and $|\delta z|<2$~cm and
all photons with energies above $100$~MeV. A first
approximation for the neutrino four-momentum $P_\nu$ is obtained
using Eq.~\ref{eq4}. 
The resulting $\Delta
m=M(\pi_{\rm s}^+K^-e^+\nu)-M(K^-e^+\nu)$ distribution for signal
events is shown in
Fig.~\ref{fig_5}a as the dashed histogram. It has a peak at $0.145$~GeV/$c^2$, 
the $D^{\ast\pm}-D^0$
mass difference, and with a full width at half maximum (FWHM) of 55~MeV/$c^2$. 

Two kinematic constraints were used to improve the resolution on the
neutrino momentum. First, the square of the invariant mass of selected
particles was calculated using $M^2(\pi_s K e\nu)=(P_\nu + P_{\pi_s K
e})^2$. For
signal events the invariant mass should equal the mass of the $D^{\ast \pm}$
meson, $m_{D^{\ast \pm}}$. Only
events with $-4$~GeV/$c^2~<M(\pi_s K e\nu)^2<36$~GeV/$c^2$ were
retained. For the selected events, the magnitude of $P_{\rm rest}$ was
rescaled by a factor $x$ requiring
\begin{equation}
M(\pi_s K e\nu)^2=(P_{\rm cms} - x\cdot P_{\rm rest})^2\equiv m_{D^{\ast \pm}}~,
\label{eq5}
\end{equation}
with $m_{D^{\ast \pm}}$ fixed to the nominal value of 2.010~GeV/$c^2$
\cite{PDG}. 
 
As a second kinematic constraint, the square of the missing
mass, $P_\nu^2$, is used. For events satisfying $-2$~GeV/$c^2~<P_\nu^2<0.5$~GeV/$c^2$, 
the
angle $\alpha$ between the direction of $\vec p_{rest}$ and the direction of the
$\pi_s K e$ system momentum was corrected in order to yield 
\begin{equation}
m_\nu^2=(E_{\rm cms}-E_{\pi_s K e}-E_{\rm rest})^2-p^2_{\pi_s K
e}-p^2_{\rm rest}-2p_{\pi_s Ke}p_{\rm rest}\cos{\alpha}\equiv 0.
\label{eq6}
\end{equation}
The angle $\alpha$ was corrected by rotating $\vec p_{rest}$
in the plane determined by the vectors $\vec p_{rest}$ and $\vec p_{\pi_s K
e}$.

The $\Delta m$ distribution obtained using the neutrino four-momentum
after the use of kinematic constraints is shown in
Fig.~\ref{fig_5}a as the solid histogram. The resolution is
significantly improved,
with the FWHM being about 7~MeV/$c^2$. It also improves at higher
values of $p^\ast(Ke)$, as illustrated in Fig.~\ref{fig_5}b. Using the simulated 
background events
it has been verified that the described neutrino reconstruction does
not induce any peaking in the background $\Delta m$ distribution. 

\begin{figure}[htb]
\includegraphics[width=0.6\textwidth]{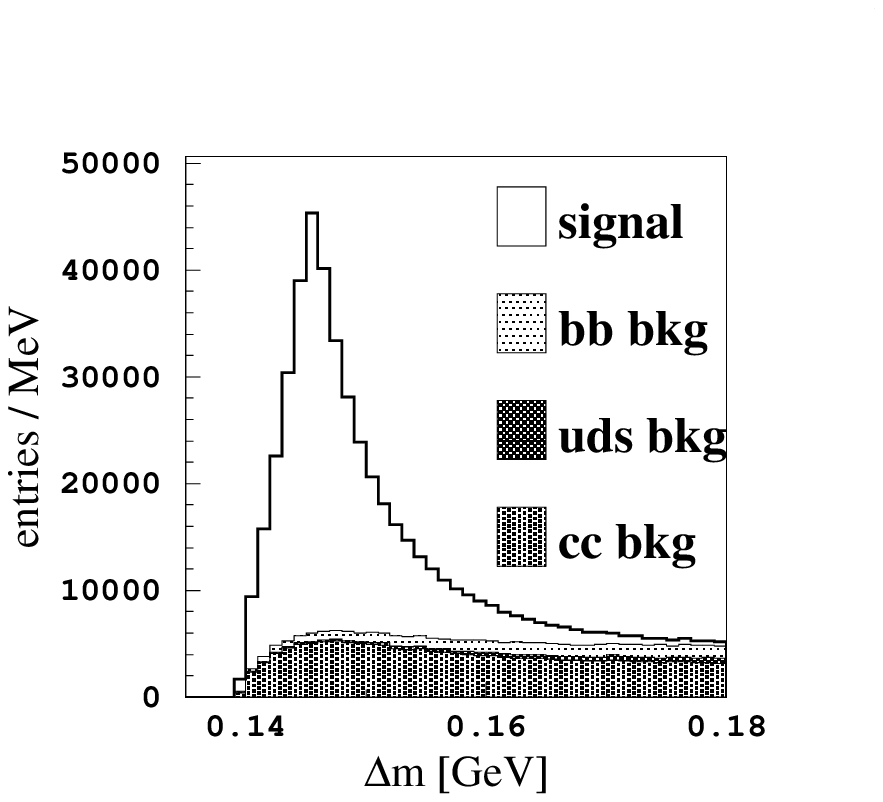}
\vspace*{-1.3cm}
\caption{Simulated distribution for $\Delta m$ with neutrino
momentum reconstruction using kinematic constraints. The open
histogram shows the contribution of the signal and the hatched
histograms show individual contributions to the background.}
\label{fig_8}
\end{figure}

The MC prediction for the reconstructed $\Delta m$
distribution is shown in Fig.\ref{fig_8}; the fraction
of signal events in the $\Delta m<0.18$~GeV/$c^2$ region is 
$f_s=0.620\pm 0.001$. The background is composed of $(75.6\pm 0.2)\%$
$c\overline{c}$ events, $(5.4\pm 0.1)\%$ light quark events and
$(18.9\pm 0.1)\%$ $B\overline{B}$ events. There is a small fraction (about 1\%) of
signal events arising from $B\overline{B}$ events. At this stage the efficiency for
reconstructing the signal with $\Delta m<0.18$~GeV/$c^2$ is found
to be $(5.4\pm 0.1)\%$. 

\section{IV Background evaluation and $\Delta m$ distribution}
%\label{sec.4}
%\setcounter{figure}{0}

\subsection{Background in the RS sample}

\begin{figure}[htb]
\includegraphics[width=1.0\textwidth]{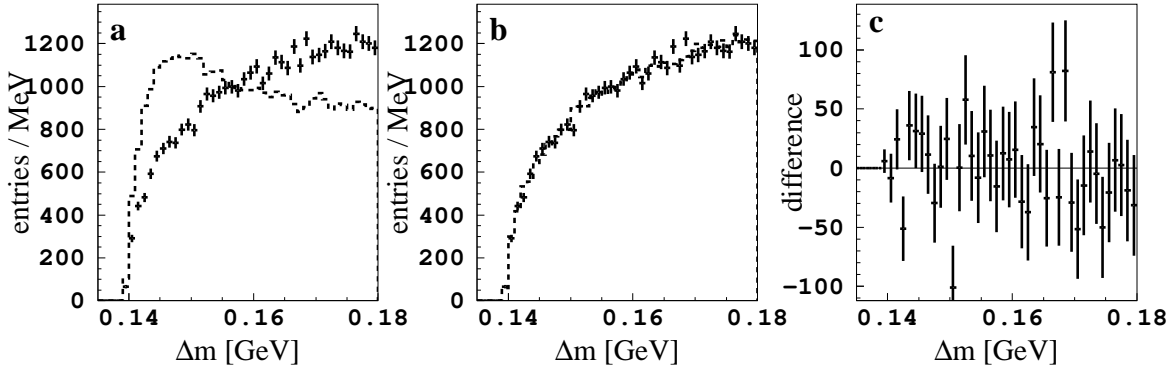}
\vspace*{-1.3cm}
\caption{a) $\Delta m$ distribution for all simulated RS background
events (dashed histogram) and for CS data events (points with error bars). Histograms
are normalized to the same number of entries. b) The same comparison after
associated signal and $D^0\to K^-\pi^+\pi^0$ decays are removed from
the simulated RS distribution. c) The difference between the  normalized $\Delta m$
distributions for CS data and the remaining simulated RS background.}
\label{fig_9}
\end{figure}

To check the MC and to avoid systematic errors arising from
a possible discrepancy with the data, the majority of the background was described using the 
data. The combinatorial, charge un-correlated background in the RS $\Delta m$
distribution can be described using
``combinatorial sign" (CS) track combinations (Table~\ref{tab1}). 
These are composed of a
slow pion candidate, $\pi^\pm_s$, and a kaon and electron of the
same charge, $K^\pm e^\pm$. Since such 
combinations cannot arise from semileptonic decays of $D$ mesons, they describe
random track combinations giving rise to a background with a $\Delta m$
value less than 0.18~GeV/$c^2$. Fig.~\ref{fig_9}a shows a comparison
of $\Delta m$ for all MC 
RS background events and CS data events; the former shows an excess of
events in the region of low $\Delta m$. 
%(see also
%Fig. \ref{fig_8}). 
This is a consequence of the charge-correlated
track combinations apart from the signal itself, present in the RS
sample.

There are additional semileptonic decays with the same charge
combination of final state charged tracks as in the $D^0\to
K^-e^+\nu$ decay. 
The most important one is $D^0\to K^-e^+\nu\pi^0$, which according to
the simulation represents \footnote{This 
includes the $D^0\to K^{\ast -}e^+\nu$ decay mode.} 8.8\% 
of all reconstructed events with $\Delta
m<0.18$~GeV/$c^2$. Other modes, like Cabibbo suppressed $D^0\to
\pi^-e^+\nu$, or $D^0\to \pi^-e^+\nu \overline{K}^0$ decays, with a charged
pion misidentified as a kaon[16], represent an additional 1\% of
simulated events in Fig.~\ref{fig_9}a. 
The charge correlation is also preserved if 
the slow
pion decays into a muon, and the latter is taken as a
slow pion candidate (0.7\% of all
reconstructed events). 

These decay channels have two important properties: they tend to peak
at low values of $\Delta m$ and secondly, the charge of the kaon or
electron candidate,
together with the charge of the slow pion from a $D^{\ast+}$ decay,
carries the same information on possible mixing as in the case of
the selected signal decay. 
In the following, these decays are treated
as a part of the signal and described as {\it associated signal}.

Apart from the processes included in the signal, there are a
number of
hadronic $D^0$ decays and decays where particle misidentification
might appear, contributing to a charge-correlated
background. 
Using measured branching ratios for the individual processes \cite{PDG},
measured and simulated average misidentification rates and estimates of
corresponding CKM elements for the DCS decays, we find 
that the most important contribution to
the charge-correlated background in the RS sample is the decay 
$D^0\to K^-\pi^+\pi^0$. This is confirmed using the MC. Contributions from 
other decays are expected to
be almost an order of magnitude smaller. The $\Delta m$ distribution and the amount of the charge
correlated background were evaluated using the MC sample. The hadronic
background accounts for 
$0.44\pm 0.01\%$ of all 
events with $\Delta m\le 0.18$~GeV/$c^2$ and is included in the CS $\Delta m$ distribution 
when extracting the yield of RS events. 

After removing the associated signal processes and the contribution from
$D^0\to K^-\pi^+\pi^0$ decays from the simulated RS background, the
agreement of the $\Delta m$ distribution with the CS data is
good. The comparison is shown in Fig.~\ref{fig_9}b and c. The $\chi^2$,
calculated from the difference of the two, has a value of 36.5 for 40 
degrees of freedom. 

\subsection{Background in the WS sample}

\begin{figure}[htb]
\includegraphics[width=1.0\textwidth]{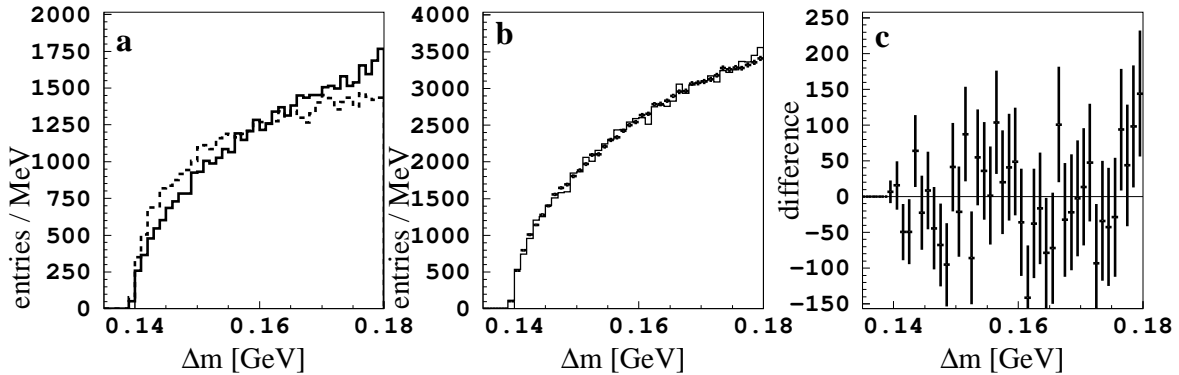}
\vspace*{-1.3cm}
\caption{a) Simulated $\Delta m$ distribution for WS (solid histogram) and RS
(dashed histogram) backgrounds. Both are normalized to the same
area. b) Comparison of simulated WS 
background without the contribution of angularly correlated background (solid histogram) 
with the background obtained using the embedded pions (points with error 
bars). c) The difference of the two (normalized) distributions
shown in the previous plot.}
\label{fig_12}
\end{figure}

While the CS events represent a good description of the combinatorial
background for the RS sample, the shape of the WS sample combinatorial
background is different (Fig.~\ref{fig_12}a). This is due to different
fractions of events, where the slow 
pion candidate is a true $\pi_s^\pm$ from a $D^{\ast\pm}$ decay and
either the kaon or electron candidate (or both, at least one being
misidentified) are from the corresponding $D^0$ decay. 
The amount of this background, which will be referred to  as the {\it angularly
correlated} background, is much smaller in the WS sample.
The rest of the background is angularly uncorrelated and represents
the majority of the background in the WS sample. Therefore WS background
events were described by 
combining a $D^0$ candidate from one event with a slow pion candidate
from another event ({\it embedded pions}). The shape of the $\Delta m$
distribution using the embedded pions agrees 
well with the MC predicted angularly uncorrelated WS background
distribution. The
comparison is shown in Fig.~\ref{fig_12}b and \ref{fig_12}c. The
$\chi^2$ of the 
difference was found to be 33.1 for 40 d.o.f. An additional $(2.04 \pm
0.05)\%$ of angularly correlated background
is not represented by using
embedded $\pi_s^\pm$.
The MC simulation shows that the largest
contribution to this background comes from $D^0\to K^-\pi^+\pi^0$,
as already established for the RS sample. 
%Other sources include $D^0\to K^-\pi^+$, $D^0\to K^-K^+$ and $D^0\to
%K^+\pi^-\pi^0$. 
The $\Delta m$ distribution for these events was obtained
from simulation and added to the background described by the embedded slow pions.

A preliminary  binned $\chi^2$ fit to the $\Delta m$ distribution of 
RS data events was performed, before any
proper decay time selection. For the 
signal and associated signal the simulated shape of $\Delta m$,
$P_s(\Delta m)$, was assumed. Background in
this sample was described using the CS data events with an additional
contribution from $D^0\to K^-\pi^+\pi^0$ events, $P_b^R(\Delta
m)$. The fraction of the 
latter was fixed in the fit to the value predicted by the MC. The data
distribution was fitted with
\begin{equation}
{\cal{P}}_R(\Delta m)={\cal{N}}_R(f_s^RP_s(\Delta m)+(1-f_s^R)P_b^R(\Delta m))~~,
\label{eq11}
\end{equation}
where the fraction of
signal events (including the associated signal) $f_s^R$ and an overall normalization 
${\cal{N}}_R$ were the
free parameters. The result of the fit was $f_s^R=0.732\pm 0.003$.  

In order to enhance the sensitivity to mixed $D^0$ decays a selection based on the proper
decay time was used, as explained in the next section. 

\section{V Proper decay time distributions}
%\label{sec.5}
%\setcounter{figure}{0}

It should be noted that in the present
analysis no attempt is made to obtain results on the mixing rate from the
fit to proper decay time distributions of $D^0$ mesons. 
The decay time is used only to
select possible mixed decays with a higher purity. To measure the
mixing rate from the ratio of WS and RS events, one needs to evaluate 
the ratio of acceptances for mixed and non-mixed
decays for decay times larger than a chosen value. The purpose of
this section is to describe the estimation of the acceptance ratio, in
which most of the systematic uncertainties cancel.

\begin{figure}[htb]
\includegraphics[width=1.0\textwidth]{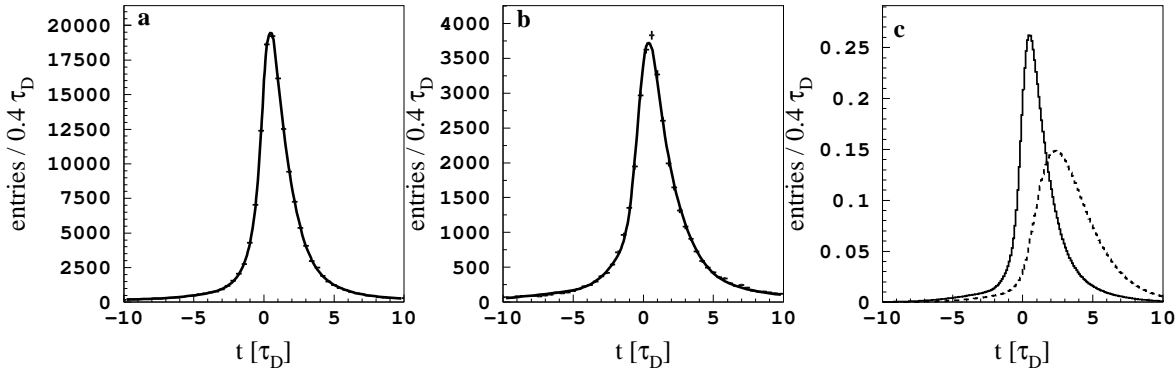}
\vspace*{-1.3cm}
\caption{a) Measured proper decay time distribution for the RS events (points
with error bars). 
The solid line shows the results of the fit described in the text.
b) Distribution of WS background events, obtained by combining a $D^0$
candidate with an embedded $\pi_s^\pm$. The solid line is again a
the result
of the fit described in the text. c) Expected decay time distribution of $D^0\to K^-e^+\nu$
decays (solid line) and $D^0\to \overline{D^0}\to K^+e^-\overline{\nu}$ (dashed line) 
calculated using the parameters obtained from the fit to the RS
event time distribution.} 
\label{fig_14}
\end{figure}

The mixed events are 
expected to have the following
proper decay time distribution \cite{time_th}:
\begin{equation}
{\cal{P}_{\rm WS}}(t^\prime)\propto t^{\prime 2}~e^{-t^\prime/\tau_{D^0}}~,
\label{eq7}
\end{equation}
with $\tau_{D^0}=(411.7\pm
2.7)\times 10^{-15}$~s \cite{PDG} denoting the
lifetime of neutral $D$ mesons. The sensitivity to mixed
events can be enhanced by selecting $D^0$ decays with a longer proper
decay time, as the background events tend to be concentrated at lower
decay times.
The decay time $t^\prime$ is evaluated using the decay length from the
$e^+e^-$ interaction point to the reconstructed $D^0$ decay vertex, and 
the measured momentum of the meson; due to the shape of the KEK-B accelerator interaction
region, which is narrowest in the vertical ($y$) direction,
the dimensionless proper decay
time $t$ is calculated as 
\begin{equation}
t={t^\prime\over\tau_{D^0}}={m_{D^0}\over p_y}~{y_{\rm vtx}-y_{\rm
IP}\over c\tau_{D^0}}~.
\label{eq8}
\end{equation}
$p_y$ is the $y$ component of the $D^0$ candidate's momentum and $y_{\rm
vtx}$ and $y_{\rm IP}$ are the $y$ coordinates of the reconstructed
$K-e$ vertex and of the interaction point, respectively. 
The observed $t$ distribution is smeared due to
the experimental resolution. Fig.~\ref{fig_14}a shows the decay time
distribution of RS data events with $\Delta m <0.18$~GeV/$c^2$. 

We fit the RS decay time distribution with a signal and a background term. The former is an 
exponential convolved with a sum of three 
Gaussians, describing the detector resolution. The background contribution is composed 
of a lifetime (exponential) and a prompt (delta function) component, convolved with the 
resolution function. The resulting value of the mean $t$ for the signal part 
is $1.093\pm 0.044$. 
A similar fit is performed on the WS 
background events (Fig.~\ref{fig_14}b), using the same ansatz as for the RS background.
From the fitted parameters we deduce the expected decay time distribution of un-mixed and 
mixed decays, 
shown in Fig.~\ref{fig_14}c. To improve the sensitivity to mixed decays, we compare the expected 
decay time distribution of mixed decays with that for background 
(Fig.\ref{fig_14}b). The optimum for the selection criterion is found
to be $t>1.5$. This requirement is imposed on both RS and WS events. By doing that,
the final ratio of WS and RS signal events depends only on the ratio of
decay time selection efficiencies for the two classes of
events, $\epsilon^{\rm unmix}/\epsilon^{\rm mix}$. 
Taking the ratio of efficiencies instead of applying the
described selection to WS events only reduces the systematic error
arising from an imperfect description of the decay time distributions.
For the ratio of efficiencies for 
$t>1.5$ we obtain $\epsilon^{\rm unmix}/\epsilon^{\rm mix}=0.420\pm
0.010$, where the error is obtained by varying all the parameters of
the fit to decay time distributions
by one standard deviation.

Table~\ref{tab4} shows the composition of the simulated sample after the proper decay 
time selection.
\begin{table}[h]
\begin{tabular}{|l|l|l|l|}
\hline\hline
Component & Fraction [\%] & Component & Fraction [\%] \\ \hline\hline
RS &   & WS & \\ \hline
signal  & $57.7\pm 0.1$  & $c\overline{c}$ bckg. & $73.6\pm 0.1$ \\ 
assoc. signal & $12.4 \pm 0.1$ &  \hspace*{0.8cm} angularly corr. bckg. & $ 2.18 \pm 0.08$\\
$c\overline{c}$ bckg. & $21.9\pm 0.1$ & $uds$ bckg. & $12.9 \pm 0.1$ \\
\hspace*{0.8cm} $D^0\to K^-\pi^+\pi^0$ & $0.55\pm 0.02$ & $B\overline{B}$ bckg. & $13.5\pm 0.1$\\ 
$uds$ bckg. & $2.63 \pm 0.06$  & & \\
$B\overline{B}$ bckg. & $4.9\pm 0.1$ & & \\ \hline\hline
\end{tabular}
\caption{Composition of the simulated selected samples. The indented contributions 
are subsamples of the above component. The quoted errors on the fractions 
are due to the MC statistics only.}
\label{tab4}
\end{table}

\section{VI Fit of $\Delta m$ distribution}
%\label{sec.6}
%\setcounter{figure}{0}
\subsection{Signal yield}

\begin{figure}[htb]
\includegraphics[width=0.6\textwidth]{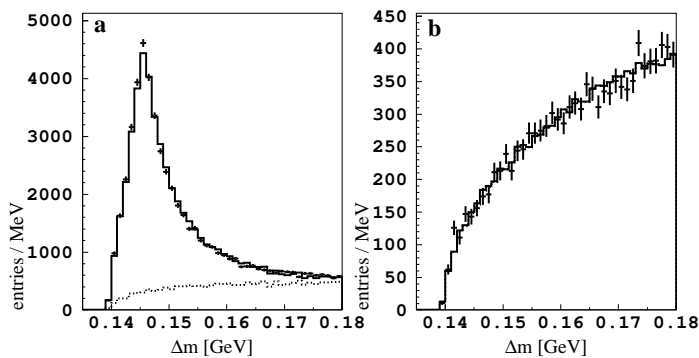}
\vspace*{-1.cm}
\caption{a) Distribution of $\Delta m$ for RS events. 
b) Distribution of $\Delta m$ for WS events. 
Data is shown
with error bars, the solid histogram shows the fit, as explained in the
text. The background contribution is shown as the dashed histogram.} 
\label{fig_16}
\end{figure}

The $\Delta m$ distributions for RS and WS events, after all selection
requirements, including $t>1.5$, are shown in Fig.~\ref{fig_16}. 

The data distributions are fitted using a binned $\chi^2$ fit. The RS
distribution is described as explained in section 4. In the fit
we minimize the value of
\begin{equation}
\chi^2=\sum_{i=1}^{N_{\rm bin}}{(N_i-{\cal{P}}_R(\Delta m_i))^2\over
\sigma^2(N_i) + \sigma^2({\cal{P}}_R(\Delta m_i))}~~.
\label{eq13}
\end{equation}
$N_i$ is the number of entries in the $i$-th bin of the RS data
distribution and ${\cal{P}}_R(\Delta m_i)$ is given by the number of
entries of MC signal, CS and simulated $D^0\to K^-\pi^+\pi^0$ events in this bin. 
Both errors, on $N_i$ and on ${\cal{P}}_R(\Delta m_i)$, 
are included in the $\chi^2$
definition.

From the
fitted values of $f_s^R$ (signal fraction) and ${\cal{N}}_R$ (overall
normalization) we obtain the number of signal non-mixed events $N_{\rm
RS}=40198\pm 329$. The quoted error includes the statistical error of
the data  as well as the error due to the limited
statistics of the expected distributions (the same holds for the fit
to the WS sample, described below). The $\chi^2$ of
the fit with 39 degrees of freedom is 69.5. The goodness-of-fit
depends strongly on the amount of the associated signal included, due
to the different widths for the $\Delta m$ distributions of
$D^0\to K^- e^+\nu$ and associated decay modes. This contribution is
treated as one of the sources of the systematic uncertainty.

For the WS sample the same description of the signal ($P_s(\Delta m)$) as for the RS
sample is used, while the background  ($P_b^W(\Delta
m)$) is described using the embedded slow pions with the addition of simulated 
angularly correlated background.
The result of the fit with 
\begin{equation}
{\cal{P}}_W(\Delta m)={\cal{N}}_W(f_s^WP_s(\Delta m)+(1-f_s^W)P_b^W(\Delta m))
\label{eq12a}
\end{equation}
is $f_s^W=(1.7\pm 6.0)\times 10^{-3}$. The quality of the
fit is good: $\chi^2=37.9$ for 39 degrees of freedom. The number 
of mixed events obtained by the fit is $N_{\rm WS}=19\pm 67$. 

\subsection{Time integrated mixing rate}

The time integrated mixing rate $r_D$ is defined as \cite{PDG}
\begin{equation}
r_D={\int_0^\infty{\cal{P}}(D^0\to\overline{D}^0\to\ell^-X^+,t)\over
\int_0^\infty{\cal{P}}(D^0\to\ell^+X^-,t)}\approx{x^2+y^2\over 2}~~,
\label{eq3}
\end{equation}
where the time dependent probabilities for un-mixed and mixed events are taken to be of 
the form $e^{-t/\tau}$ and $t^2e^{-t/\tau}$, respectively. The dimensionless
mixing parameters $x$ and $y$ are equal to $x=\Delta M/\Gamma$ and $y=\Delta\Gamma/2\Gamma$, where
$\Delta M$ is the mass difference of the two neutral D meson mass
eigenstates, $\Delta\Gamma$ is the difference of the decay widths and 
$\Gamma$ is the average decay width of the two.

The parameter $r_D$ is
obtained from the fitted WS and RS yields,
\begin{equation}
r_D={N_{\rm WS}\over N_{\rm RS}}\cdot {\epsilon^{\rm unmix}\over
\epsilon^{\rm mix}}~~,
\label{eq14}
\end{equation}
with the ratio of efficiencies for decay time selection given in
Section 5. The result is
\begin{equation}
r_D=(0.20\pm 0.70)\times 10^{-3}~~,
\label{eq15}
\end{equation}
where the quoted uncertainty is statistical only. 
%Since the result is
%compatible with zero we calculate the 90\% C.L. upper limit for $r_D$:
%\begin{equation}
%r_D\le 1.10\times 10^{-3}~~{\rm at~90\%~C.L.}
%\label{eq16}
%\end{equation}

%The above upper limit is calculated from the statistical
%uncertainty alone, assuming a Gaussian distribution for the error. 

\section{VII Systematic uncertainties}
%\label{sec.7}
%\setcounter{figure}{0}

\subsection{Estimation of systematic errors}

The value of the time integrated mixing rate is obtained using the ratio of WS and RS 
$D^0$ decays. 
Since the kinematics of these decays is the same, the selection efficiencies, excluding the 
acceptance 
of the proper decay time selection, are expected to be the same for 
both types of decays and hence cancel in the final expression (\ref{eq14}). 
Using the MC sample of events we find the selection efficiency ratio to be $0.97\pm 0.05$, 
compatible
with this expectation.

\begin{figure}[htb]
\includegraphics[width=0.35\textwidth]{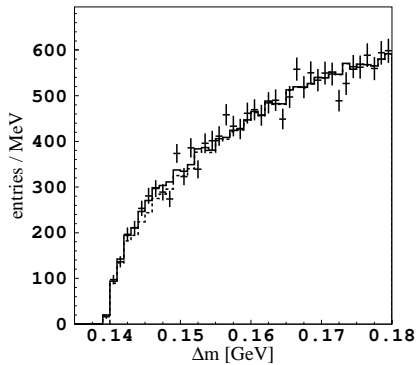}
\vspace*{-.8cm}
\caption{Distribution of $\Delta m$ for simulated events including
mixed $D$ mesons (points with error bars), the fit (solid line) and the background
(dashed line).}
\label{mixed}
\end{figure}

The influence of the kinematic constraints in the neutrino
reconstruction on the $\Delta m$ distribution of background has been
verified using the MC simulation and no artificial peaking has been
observed. The fit to the RS sample $\Delta m$ distribution correctly
reproduces the number of signal events. For the simulated value of
$N_{\rm RS}^{\rm in}=59561$ we obtain $N_{\rm RS}^{\rm out}=59193\pm
498$. Although the numbers of RS events agree within the statistical
error of the fit, the remaining relative difference is propagated to
an absolute error on $r_D$ of $0.001\times 10^{-3}$ and is added to
the total systematic error. The reconstruction of WS events has also
been verified using the simulated sample. To a sample of $2.93\times
10^{6}$ non-mixed events, roughly corresponding to an integrated
luminosity of 180~fb$^{-1}$, we added 6000 mixed $D$ meson decays,
obtaining a sample of $D^0$ decays with an input value of $r_D^{GEN} =
2.05\times 10^{-3}$. The fit (Fig.~\ref{mixed}) with the same method
as used for the data yields $N_{\rm WS}^{\rm out} = 234 \pm 85$ and
$N_{RS} = 48459 \pm 420$. The former can be compared to the simulated
value of $N_{\rm WS}^{\rm in} = 255$. The fitted numbers of signal
events result in $r_D = (2.0 \pm 0.7)\times 10^{-3}$, which very well
reproduces the input value. Here also the relative difference between
the simulated and reconstructed $N_{\rm WS}$ is taken into account in
the final systematic error. 

The error on the ratio of decay time selection efficiencies
$\epsilon^{\rm unmix}/\epsilon^{\rm mix}$ is obtained by varying the
fitted parameters of the RS and WS sample proper decay time
distributions and leads to an additional $0.005\times 10^{-3}$ error on
$r_D$. We evaluated the same efficiency ratio using the MC sample and obtained
$\epsilon^{\rm unmix}/\epsilon^{\rm mix}=0.400\pm 0.012$,  which is
consistent with the value obtained from the proper decay time fit. 
 The proper decay time
selection was additionally verified by performing the fit to RS and WS
$\Delta m$ distributions using several different decay time
requirements, $t>1.2$ and $t>1.7$. The resulting values of $r_D$ are
$(0.30\pm 0.70)\times 10^{-3}$ and $(0.21\pm 0.70)\times 10^{-3}$,
respectively. 
%While the measured values (including the default one for
%$t>1.5$) agree within the statistical errors of the fit\footnote{It
%should be noted that the statistical errors of individual results are
%correlated.}, 
We take the largest difference as a systematic error due
to an imperfect description of the decay time distribution.

The next possible source of systematic uncertainty arises from  MC samples used to fit the 
$\Delta m$ distributions. Using the measured branching fractions and corresponding errors \cite{PDG}
of decay channels 
included in the associated signal, and their relative contributions to the selected sample 
as obtained by MC, we estimate 
that the amount of this component in the fit needs to be varied by 40\%. The fit was repeated 
and the difference in the obtained values of $r_D$ is found to be $0.015\times 10^{-3}$.

The fraction of the angularly correlated WS background is estimated from MC to 
be $(2.04\pm 0.05)\%$. Since the majority 
of these events arise from $D^0\to K^-\pi^+\pi^0$ decays, we repeated the WS fit by changing
the amount of this background in accordance with the relative error on
$B(D^0\to K^-\pi^+\pi^0)$ \cite{PDG}. 
The variation of $N_{\rm WS}$ resulted in a $0.05\times 10^{-3}$ error on $r_D$.

The systematic uncertainties are summarized in Table~\ref{tab3}. By summing the individual
contributions in quadrature, we obtain the total systematic error of $0.11\times 10^{-3}$. It 
is clear that the
uncertainty of the result on the time integrated mixing rate is limited by the statistics of 
the data.

\begin{table}[h]
\begin{tabular}{|c|c|c|}
\hline\hline
Source & Estimated by & Resulting $r_D$ uncertainty $[\times 10^3]$\\ \hline\hline
$\nu$ reconstruction & $|N_{\rm RS}^{\rm in}-N_{\rm RS}^{\rm out}|/N_{\rm RS}^{\rm in}$ & $\pm 0.001$\\ 
                     & $|N_{\rm WS}^{\rm in}-N_{\rm WS}^{\rm out}|/N_{\rm WS}^{\rm in}$ & $\pm 0.016$\\ 
\hline
ratio of $t$ selection & $\pm 1\sigma$ of $t$ distribution fit parameters & $\pm 0.005$ \\
efficiencies           & using different $t$ selection values         & $\pm 0.100$ \\ \hline
amount of associated signal & $\pm 40\%$ & $\pm 0.015$ \\ \hline
amount of charge correlated & $\pm 7\%$ & $\pm 0.050$ \\
WS backg. ($D^0\to K^-\pi^+\pi^0$) & & \\ \hline\hline
Total      & & $\pm 0.11$ \\ \hline\hline
\end{tabular}
\caption{Summary of systematic measurement uncertainties. The
evaluation of individual contributions are described in the text.}
\label{tab3}
\end{table}

\subsection{Checks of the analysis method}

Apart from the checks on the quality of background description described in 
previous sections, several other tests of
the analysis method were performed. 

The fits to the RS $\Delta m$ distribution were done individually for
different data taking periods. The resulting signal fractions 
show no significant deviation from the overall result
used in the analysis. 

In order to check for a possible bias in the result, we performed a
fit to the $\Delta m$ distribution of WS events using an inverted
decay time selection, i.e. fitting the events with $t<1.5$. A possible 
mis-assignment of the charge of the slow pion candidate, for example, would 
appear as a peak in this sample and could bias the fitted number of WS events. 
Since $t<1.5$ is the region where the sensitivity to mixed events is reduced, we
expect no such contribution and any positive signal would be an indication of 
a peaking background behavior. 
The fit was done in the same manner as the fit
to $t>1.5$ distribution and the resulting number of mixed events is $N_{\rm
WS}=-77\pm 88$, in accordance with the expectation. The fitted number of non-mixed decays
is $N_{\rm RS}=82943\pm 456$, which taking into account the corresponding ratio of 
decay time selection efficiencies gives $r_D=(-3.1\pm 3.6)\times 10^{-3}$.

By repeating the fit procedure without the decay time selection, we
obtain $N_{\rm WS}=-48\pm 111$ and $N_{\rm RS}=123054\pm 562$. 
While the central value of $r_D=(-0.39\pm 0.90)\times 10^{-3}$ 
is compatible with the null 
result, as for the events with $t>1.5$, the
statistical error is increased, confirming the usefulness of the decay time
selection. 

To test the bias as well as the obtained statistical error of $N_{\rm
WS}$ we divided the 
available simulated WS sample into two parts. The corresponding
$\Delta m$ distributions were then fitted one to another. The result
of the fit is $N_{\rm WS}=-66\pm 62$. The statistical error of the
``background'' distribution was not included in this fit. Afterward we
independently varied the contents 
of each bin within the statistical error and repeated the fit many
times. 
The distribution of resulting $N_{\rm WS}$
is centered at the starting value (due to a
statistical fluctuation when arbitrarily dividing the MC sample into
two parts) with a width of $61.9\pm 0.5$. The latter represents the expected
statistical error of the fit on a sample with about the same
statistics as the data itself. The expected error is in agreement with the
fit to the data 
when not taking into consideration the statistical error of the
expected distributions ($\pm 64$).

\section{VII Conclusions}
%\label{sec.8}
%\setcounter{figure}{0}

We have searched for $D^0-\overline{D^0}$ mixing in semileptonic 
$D$ meson decays. The resulting time integrated mixing rate is 
\begin{equation}
r_D=(0.20 \pm 0.70 \pm 0.11)\times 10^{-3}~~,
\label{eq_res1}
\end{equation}
where the first error represents the statistical and the second the
systematic uncertainty.
 
Since no significant signal for the WS decays $D^0\to K^+e^-\nu$ is
observed, we put an upper limit on the time integrated mixing rate.
We combine statistical and systematic errors in quadrature and asume a
Gaussian distribution of the resulting total error;
 using the Feldman Cousins
approach\cite{FeldmanCousins} to estimate the upper limit in the
vicinity of the physical boundary ($r_D\ge 0$) we find
\begin{equation}
r_D\le 1.4\times 10^{-3}~~{\rm at~90\%~C.L.}
\label{eq_res}
\end{equation}

The limit obtained is more restrictive than the existing world average 
$\Gamma(K^+\ell^-\overline{\nu})/\Gamma(K^-\ell^+\nu)<5\times 10^{-3}$\cite{PDG}. 
Using expression (\ref{eq3}) and the value of 
$\Delta\Gamma /\Gamma=2y=0.016\pm0.010$\cite{PDG}, the above result can be expressed 
as
\begin{equation}
\Delta M\le 13\times 10^{10}~\hbar s^{-1} ~~{\rm at~90\%~C.L.}
\label{eq_dm}
\end{equation}

\section*{Acknowledgments}
%***** Acknowledgments *****
% Please paste this acknowledgement into your latex file. 
 %----------- Long version, for most papers ----------- 
We thank the KEKB group for the excellent operation of the
accelerator, the KEK Cryogenics group for the efficient
operation of the solenoid, and the KEK computer group and
the National Institute of Informatics for valuable computing
and Super-SINET network support. We acknowledge support from
the Ministry of Education, Culture, Sports, Science, and
Technology of Japan and the Japan Society for the Promotion
of Science; the Australian Research Council and the
Australian Department of Education, Science and Training;
the National Science Foundation of China under contract
No.~10175071; the Department of Science and Technology of
India; the BK21 program of the Ministry of Education of
Korea and the CHEP SRC program of the Korea Science and
Engineering Foundation; the Polish State Committee for
Scientific Research under contract No.~2P03B 01324; the
Ministry of Science and Technology of the Russian
Federation; the Ministry of Education, Science and Sport of
the Republic of Slovenia; the National Science Council and
the Ministry of Education of Taiwan; and the U.S.\
Department of Energy.

%-------- Short version, if necessary, for PRL -----------
% currently commented out
%We thank the KEKB group for the excellent operation of the
%accelerator, the KEK Cryogenics group for the efficient
%operation of the solenoid, and the KEK computer group and
%the NII for valuable computing and Super-SINET network
%support.  We acknowledge support from MEXT and JSPS (Japan);
%ARC and DEST (Australia); NSFC (contract No.~10175071,
%China); DST (India); the BK21 program of MOEHRD and the CHEP
%SRC program of KOSEF (Korea); KBN (contract No.~2P03B 01324,
%Poland); MIST (Russia); MESS (Slovenia); NSC and MOE
%(Taiwan); and DOE (USA).

\end{document}

%% file: author-conf2004.tex
%%% Paper:    
%%% Journal:  summer 2004 conference papers (PRL format)
%%% Contacts: 
%%% Last revised on July 14, 2004 16:40:00 EDT
%%% Non-responding authors or those who said NO are commented out.
%%% ====================================================================
%%% Click the RELOAD button on your web browser to see the updated file.
%%% ====================================================================
%%% Use \input{author} to insert this material into your latex file.
%%%%% Force institutions to appear in alphabetical order when typeset.
\affiliation{Aomori University, Aomori}
\affiliation{Budker Institute of Nuclear Physics, Novosibirsk}
\affiliation{Chiba University, Chiba}
\affiliation{Chonnam National University, Kwangju}
\affiliation{Chuo University, Tokyo}
\affiliation{University of Cincinnati, Cincinnati, Ohio 45221}
\affiliation{University of Frankfurt, Frankfurt}
\affiliation{Gyeongsang National University, Chinju}
\affiliation{University of Hawaii, Honolulu, Hawaii 96822}
\affiliation{High Energy Accelerator Research Organization (KEK), Tsukuba}
\affiliation{Hiroshima Institute of Technology, Hiroshima}
\affiliation{Institute of High Energy Physics, Chinese Academy of Sciences, Beijing}
\affiliation{Institute of High Energy Physics, Vienna}
\affiliation{Institute for Theoretical and Experimental Physics, Moscow}
\affiliation{J. Stefan Institute, Ljubljana}
\affiliation{Kanagawa University, Yokohama}
\affiliation{Korea University, Seoul}
\affiliation{Kyoto University, Kyoto}
\affiliation{Kyungpook National University, Taegu}
\affiliation{Swiss Federal Institute of Technology of Lausanne, EPFL, Lausanne}
\affiliation{University of Ljubljana, Ljubljana}
\affiliation{University of Maribor, Maribor}
\affiliation{University of Melbourne, Victoria}
\affiliation{Nagoya University, Nagoya}
\affiliation{Nara Women's University, Nara}
\affiliation{National Central University, Chung-li}
\affiliation{National Kaohsiung Normal University, Kaohsiung}
\affiliation{National United University, Miao Li}
\affiliation{Department of Physics, National Taiwan University, Taipei}
\affiliation{H. Niewodniczanski Institute of Nuclear Physics, Krakow}
\affiliation{Nihon Dental College, Niigata}
\affiliation{Niigata University, Niigata}
\affiliation{Osaka City University, Osaka}
\affiliation{Osaka University, Osaka}
\affiliation{Panjab University, Chandigarh}
\affiliation{Peking University, Beijing}
\affiliation{Princeton University, Princeton, New Jersey 08545}
\affiliation{RIKEN BNL Research Center, Upton, New York 11973}
\affiliation{Saga University, Saga}
\affiliation{University of Science and Technology of China, Hefei}
\affiliation{Seoul National University, Seoul}
\affiliation{Sungkyunkwan University, Suwon}
\affiliation{University of Sydney, Sydney NSW}
\affiliation{Tata Institute of Fundamental Research, Bombay}
\affiliation{Toho University, Funabashi}
\affiliation{Tohoku Gakuin University, Tagajo}
\affiliation{Tohoku University, Sendai}
\affiliation{Department of Physics, University of Tokyo, Tokyo}
\affiliation{Tokyo Institute of Technology, Tokyo}
\affiliation{Tokyo Metropolitan University, Tokyo}
\affiliation{Tokyo University of Agriculture and Technology, Tokyo}
\affiliation{Toyama National College of Maritime Technology, Toyama}
\affiliation{University of Tsukuba, Tsukuba}
\affiliation{Utkal University, Bhubaneswer}
\affiliation{Virginia Polytechnic Institute and State University, Blacksburg, Virginia 24061}
\affiliation{Yonsei University, Seoul}
  \author{K.~Abe}\affiliation{High Energy Accelerator Research Organization (KEK), Tsukuba} % KEK
  \author{K.~Abe}\affiliation{Tohoku Gakuin University, Tagajo} % TohokuGakuin
  \author{N.~Abe}\affiliation{Tokyo Institute of Technology, Tokyo} % TIT
  \author{I.~Adachi}\affiliation{High Energy Accelerator Research Organization (KEK), Tsukuba} % KEK
  \author{H.~Aihara}\affiliation{Department of Physics, University of Tokyo, Tokyo} % Tokyo
  \author{M.~Akatsu}\affiliation{Nagoya University, Nagoya} % Nagoya
  \author{Y.~Asano}\affiliation{University of Tsukuba, Tsukuba} % Tsukuba
  \author{T.~Aso}\affiliation{Toyama National College of Maritime Technology, Toyama} % Toyama
  \author{V.~Aulchenko}\affiliation{Budker Institute of Nuclear Physics, Novosibirsk} % BINP
  \author{T.~Aushev}\affiliation{Institute for Theoretical and Experimental Physics, Moscow} % ITEP
% ...
% Lausanne
\collaboration{The Belle Collaboration}